%% file: main.tex
\documentclass[twocolumn,table,xcolor]{aastex63}
\usepackage{xcolor}
\usepackage{dirtytalk}
\usepackage{amsmath}
\usepackage{array}
\usepackage{float}
\usepackage{enumitem}
\usepackage{CJKutf8}
\usepackage{epsfig} 
\usepackage{amsmath}
\usepackage{multirow}
\newcommand{\chinesename}{{\begin{CJK}{UTF8}{gbsn}(王加冕)\end{CJK}}}
\newcommand{\kepler}{{\em Kepler} \ }
\newcommand{\hundredearths}{{\em {100} \ Earths \ Survey}}

\newcommand{\ms}{\mbox{m s$^{-1}$}}
\newcommand{\cms}{\mbox{cm s$^{-1}$}}

\newcommand{\kms}{\mbox{km s$^{-1}$}}

\newcommand{\msun}{M$_{\sun}$}
\newcommand{\rsun}{R$_{\sun}$}

\newcommand{\mearth}{M$_{\earth}$}

\newcommand{\msini}{$M \sin i$}
\newcommand{\vsini}{$v \sin i$}

\newcommand{\teff}{${\rm T_{eff}}$}
\newcommand{\sn}{$\rm S/N$}

\newcommand{\feh}{\ensuremath{[\mbox{Fe}/\mbox{H}]}}
\newcommand{\rphk}{\ensuremath{R'_{\mbox{\scriptsize HK}}}}

\newcommand{\mv}{\ensuremath{M_{\mbox{\scriptsize V}}}}

\received{May 8, 2020}
\revised{June 1, 2020}
\accepted{June 2, 2020}
\submitjournal{Astronomical Journal}

\shorttitle{EPRV Benchmark}
\shortauthors{Brewer et al.}

\graphicspath{{./}{Figures/}}

\begin{document}

\title{EXPRES I. HD~3651 an Ideal RV Benchmark}

\input{authors}



\begin{abstract}
The next generation of exoplanet-hunting spectrographs should deliver up to an order of magnitude improvement in radial velocity precision over the standard 1 \ms\ state of the art. This advance is critical for enabling the detection of Earth-mass planets around Sun-like stars. New calibration techniques such as laser frequency combs and stabilized etalons ensure that the instrumental stability is well characterized.  However, additional sources of error include stellar noise, undetected short-period planets, and telluric contamination. To understand and ultimately mitigate error sources, the contributing terms in the error budget must be isolated to the greatest extent possible. 
Here, we introduce a new high cadence radial velocity program, the EXPRES \hundredearths, which aims to identify rocky planets around bright, nearby G and K dwarfs. We also present a benchmark case: the 62-d orbit of a Saturn-mass planet orbiting the chromospherically quiet star, HD~3651. The combination of high eccentricity (0.6) and a moderately long orbital period, ensures significant dynamical clearing of any inner planets. Our Keplerian model for this planetary orbit has a residual RMS of 58 \cms\ over a $\sim 6$ month time baseline. By eliminating significant contributors to the radial velocity error budget, HD~3651 serves as a standard for evaluating the long term precision of extreme precision radial velocity (EPRV) programs.

\end{abstract}



\section{Introduction} \label{sec:intro}
Following the early detections of gas giant planets around Sun-like stars, radial velocity (RV) surveys saw a steady stream of discoveries, punctuated by regular improvements in instrumental precision.  With the introduction of the environmentally stabilized HARPS spectrograph in 2003 \citep{mayor_2003}, single measurement RV precision reached $\sim 1$ \ms.  Multi-decade campaigns continued to push to lower mass planets and longer period orbits, but the state of the art RV precision \citep{fischer_eprv_2016} has remained at this level for more than a decade. 

The long-term nature of RV surveys has enabled the collection of thousands of high-fidelity measurements for a range of spectral types. Many of these stars are radial velocity standards without detected planets. The RV RMS scatter of standard stars or the residuals after fitting a simple Keplerian model has been used to evaluate RV measurement precision. The quietest stars have shown a scatter of just under $\sim 2$ \ms\ \citep{Brems:2019_rv_jitter,Soubiran:2018_rv_standards,Huang:2018_rv_standards}.  However, it is unclear how this RV scatter is apportioned between astrophysical, instrumental, and analysis error sources.

The \kepler and K2 transit surveys have demonstrated that nearly every star hosts at least one planet \citep{Burke:2015_occur,Hsu:2018_occur}. The most commonly detected transiting planet has a radius between $1 - 4 R_\oplus$ \citep{Burke:2015_occur} and many of the transiting planet architectures contain tightly packed systems of small planets \citep{winn_occurrence_2015} that would produce short-period, low-amplitude reflex velocities in the host stars. Analogs of the \kepler rocky planets and compact multi-planet systems are largely missing from RV surveys. This implies that at least some of the RV scatter in standard stars is likely caused by undetected low-mass planets.  Both improved RV precision and higher observing cadence are required to tease out these signals. Since a reliable sample of ``stars without planets'' does not exist, a new type of standard star is needed to evaluate improvements in RV precision.

The Extreme Precision Spectrometer (EXPRES) is one of the first in a new generation of Extreme Precision Radial Velocity (EPRV) instruments delivering high-fidelity data with the goal of disentangling photospheric velocities from Keplerian velocities. EXPRES is located at the Lowell Discovery Telescope \citep[LDT,][]{Levine:2012_dct_commision,DeGroff:2014_dct_performance}. The instrumental stability is at least 10 \cms\ \citep{Blackman_2020} with single measurement uncertainties of about $30$ \cms\ in spectra with \sn\ $\sim 250$ per pixel at 550~nm \citep{Petersburg_2020}. The primary science mission for EXPRES is the \hundredearths\ to identify low-mass planets in habitable zone orbits around Sun-like stars.  The combination of high-precision measurements and high observing cadence will enable the detection of planets that were commonly found with the \kepler mission, but that have been missed in previous radial velocity surveys.

In this paper, we highlight our radial velocity data for \object{HD~3651~b} as a way to evaluate the long term, on-sky precision of EPRV instruments. The high eccentricity of this planetary orbit dynamically clears out most simulated test particles within and slightly outside its orbit, as expected from stability theory \citep{Gladman1993}. Similar orbital parameters are known for only a handful of detected exoplanets. HD~3651 is especially well-suited as a standard star for demonstrating RV precision because this bright star is accessible by all current EPRV instruments.

\section{EXPRES}

EXPRES was fully commissioned in February 2019 and has been used to collect science observations for the \hundredearths\ since that time. The high resolution optical spectrograph is a fiber fed echelle design \citep{jurgenson_expres_2016} with double scrambling and active agitation \citep{petersburg_modal_2018}. It covers 3800-7800 \AA and has a median resolving power of $R = 137,500$. EXPRES is located in a vibration isolated vacuum enclosure, in a temperature controlled room.  The front end module contains an atmospheric dispersion compensator (ADC) and a fast tip-tilt system keeps the star focused on the 0\farcs9 octagonal input fiber. The overall seeing-dependent throughput is $\sim 8\% - 15\%$ \citep{Blackman_2020}.  Wavelength calibration is carried out with a Menlo Systems laser frequency comb \citep[LFC,][]{probst_relative_2016}, and we have demonstrated an instrumental precision better than 10 cm/s \citep{Blackman_2020}. A chromatic exposure meter picks off 2\% of the light to monitor photon arrival times \citep{blackman_accounting_2017,blackman_measured_2019}. The current single measurement precision is 30 \cms\ at \sn\ of 250 per pixel \citep{Petersburg_2020}, meeting the spectrograph design goals.  Further work is underway to mitigate the impact of photospheric noise on the radial velocities. All instrument adjustments and observations are handled through a python-based messaging server and associated database with a web front end, enabling high cadence observations with minimal overhead.

\subsection{The Science Goals}
The primary mission for EXPRES is the \hundredearths, which will search for low-mass planets with orbital radii stretching out to the habitable zones of Sun-like stars. These discoveries will reach a new parameter space for RV surveys by detecting planets that are more likely to have habitable conditions orbiting nearby stars. Furthermore, the discovery of lower mass planets will help to reconcile the currently discrepant results between transit and radial velocity searches. The science goals of EXPRES will be achieved by combining high-precision, high-cadence observations with a long term monitoring program at the LDT. EXPRES can also be used for followup of transiting planets around bright stars discovered with the TESS mission \citep{Ricker_2015} and the instrument is being used to characterize exoplanet atmospheres with high-dispersion spectroscopy \citep{Hoeijmakers_2020}.

\subsection{Stellar Targets for the 100 Earths Survey} 

The primary targets for the \hundredearths\ include 66 G and K dwarfs distributed over the northern sky; most of these stars are brighter than $V \sim 7$ (Table \ref{tab:targets}) and were selected to be chromospherically quiet, without detected gas giant planets.  A few stars with high chromospheric activity or known planets were also included; these stars serve as benchmarks to evaluate the on-sky performance of the program.  They allow us to search for low-mass planetary companions to known gas giants, and they provide excellent data sets for developing statistical mitigation strategies for stellar activity. 

\input{Tables/EXPRES_targets.tex}

\subsection{Exposure Times}

The detection of low-mass planets orbiting in the stellar habitable zone requires high \sn\ observations. \citet{Petersburg_2020} show that the EXPRES single-measurement RV errors decrease with increasing \sn; dropping from errors of about 90 \cms\ for \sn\ $\sim 100$ (per pixel at 550 nm) to about 30 \cms\ for \sn\ of $250$. 
At \sn\ $> 250$, the curve flattens with minimal gains in RV measurement precision.  To minimize spurious velocity shifts due to charge transfer inefficiency \citep{Blackman_2020}, all stars are observed at a consistent \sn, and based on our analysis of the dependence of RV error on \sn\ for our target stars, we have set this to be \sn\ $= 250$ per pixel at 550 nm. This \sn\ is well below the saturation of the detector and with the 4-pixel line spread function (LSF) of EXPRES, this yields \sn\ $= 500$ per resolution element for each exposure.  To ensure that we reach a perfectly ``baked'' level in our spectra, our chromatic exposure meter picks off a fraction of the light entering the spectrograph and counts photons in the V-band.  The exposure meter counts have been calibrated to measured \sn\ in the extracted spectra and the exposure meter terminates the exposure when one of the following conditions is met: either a \sn\ of 250 has been reached or 20 minutes have elapsed (whichever comes first). The twenty minute limit for exposure times is set to minimize errors in the chromatic barycentric correction \citep{blackman_measured_2019}. For most stars, the resulting exposure times span or exceed the peak period of p-mode oscillations \citep{chaplin_filtering_2019}. For brighter stars, additional observations are obtained to average over p-mode oscillations and for very faint stars additional observations are needed to reach our desired \sn. 

\subsection{Cadence}

We initially started the \hundredearths\ with four consecutive observations for every target. After we had accumulated a 6-month data set, we randomly removed one of the 4 observations and found that when we refit our data there was almost no increase in the residual velocity RMS. With our 4-pixel LSF, three consecutive observations of spectra still yields a \sn\ of $250 \times sqrt(4) \times sqrt(3) = 866$ per resolution element in the nightly binned data, and reducing the number of consecutive observations from four to three has the important benefit that more targets can be covered each night.
Therefore, every target on the \hundredearths\ is now observed three times per night, each time it has been scheduled, weather permitting.  Under the assumption of white noise, three exposures should improve the single measurement precision of 30 \cms\ to a nightly binned measurement precision of $\sim 17$ \cms.  To track any small instrumental drifts in the wavelength solution during the night, science observations are interspersed with LFC frames every 15-30 minutes. The 10-second LFC observations have a readout time of 27 seconds and are generally taken during telescope slew times, so very little time (no more than about 10 minutes per full night) is lost taking these calibrations. The \hundredearths\ is currently allocated up to 70 nights/year on the 4.3 meter Lowell Discovery Telescope (LDT). Most of these nights are scheduled as half or quarter nights to maximize cadence on the target stars.

\subsection{Data Reduction and Analysis}

The EXPRES analysis pipeline uses a flat-relative optimal extraction algorithm \citep{Petersburg_2020}.  Each night, 30 dark and 30 science flat images are taken and used to reduce and extract the science frames. Order tracing is accomplished using the reduced science flats and after a scattered light model is removed, the flat-relative optimal extraction is performed. Wavelength solutions are interpolated from bracketed LFC images taken throughout the night and a nightly exposure of a Thorium Argon calibration lamp is used as a calibration reference to initiate the LFC wavelength solution.  The chromatic, flux-weighted midpoint time is calculated from the exposure meter data stored in a FITS header table with each spectrum. After telluric line identification and masking is done using SELENITE \citep{leet_tellurics_2019}, absolute radial velocities are derived using a forward model \citep{Petersburg_2020}.

\subsection{Validating On-Sky Precision}

EXPRES has met its design specifications \citep{Blackman_2020}.  \citet{Petersburg_2020} showed that Keplerian fitting of 47 observations for 51~Peg~b yielded orbital parameters consistent with literature values with an RMS scatter in the residual velocities of $88$\ \cms. This indicates that there is residual RV scatter from some combination of the stellar photosphere, the instrument, and our analysis pipeline. It is common, especially when testing new instruments or analyses, to choose a `standard' star to evaluate the RV performance.  However, an improvement in precision can mean that previously well-characterized stars may reveal surprises.  Given that we expect low-amplitude, high-frequency signals (i.e. small rocky planets, compact systems) around a significant fraction of stars \citep{winn_occurrence_2015}, some of radial velocity standard stars may harbor planets in the RV `noise'. It would be helpful to have even one case where we could be sure that additional planets were not contributing to the residual RV scatter. 

\section{HD~3651: An EPRV Calibrator}

To evaluate our measurement precision, we wanted to rule out contamination from low-amplitude, short-period planets.  Instead of focusing on `RV quiet' stars, we selected a star with a known planet in a very eccentric and moderately long period orbit.  Our simulations show that short-period planets should not be able to survive (Section \ref{sec:3651_interior_planets}), leaving the exoplanetary system free of RV scatter from undetected exoplanets.

HD~3651 is an old nearby K dwarf with stellar parameters summarized in Table \ref{tab:pars}. The star hosts a Saturn-mass planet (\msini $\sim 70 M_\oplus$) on a $\sim 62$ day orbit with an eccentricity of $0.61$.  Since its discovery \citep{fischer_2003}, additional observations have been obtained using Keck HIRES \citep{butler_hires_2017}.  Using 161 archival observations taken over 17 years, we fit a single planet model to the data with an RMS scatter to the residuals of 3.4 m/s (Table \ref{tab:3651_orb}). This is a few times larger than the single measurement precision of HIRES and close to the low end of the distribution of RV RMS scatter for Keck HIRES \citep{fischer_eprv_2016}.  The low activity of the star (\rphk $ = -5.01$) makes this an ideal target with low intrinsic stellar jitter \citep{isaacson_chromospheric_2010}. The RMS in the HIRES RV residuals suggests that instrumental or analytical uncertainties may dominate the error budget.  For comparison, \citet{Cosentino:2014_3651_harpsn} found that a single planet fit to two years of data using the HARPS-N spectrograph gave an RMS scatter of 1.82 \ms\ for HD~3651. 

\input{Tables/stellar_pars}

\begin{figure}
    \centering
    \includegraphics[width=\linewidth]{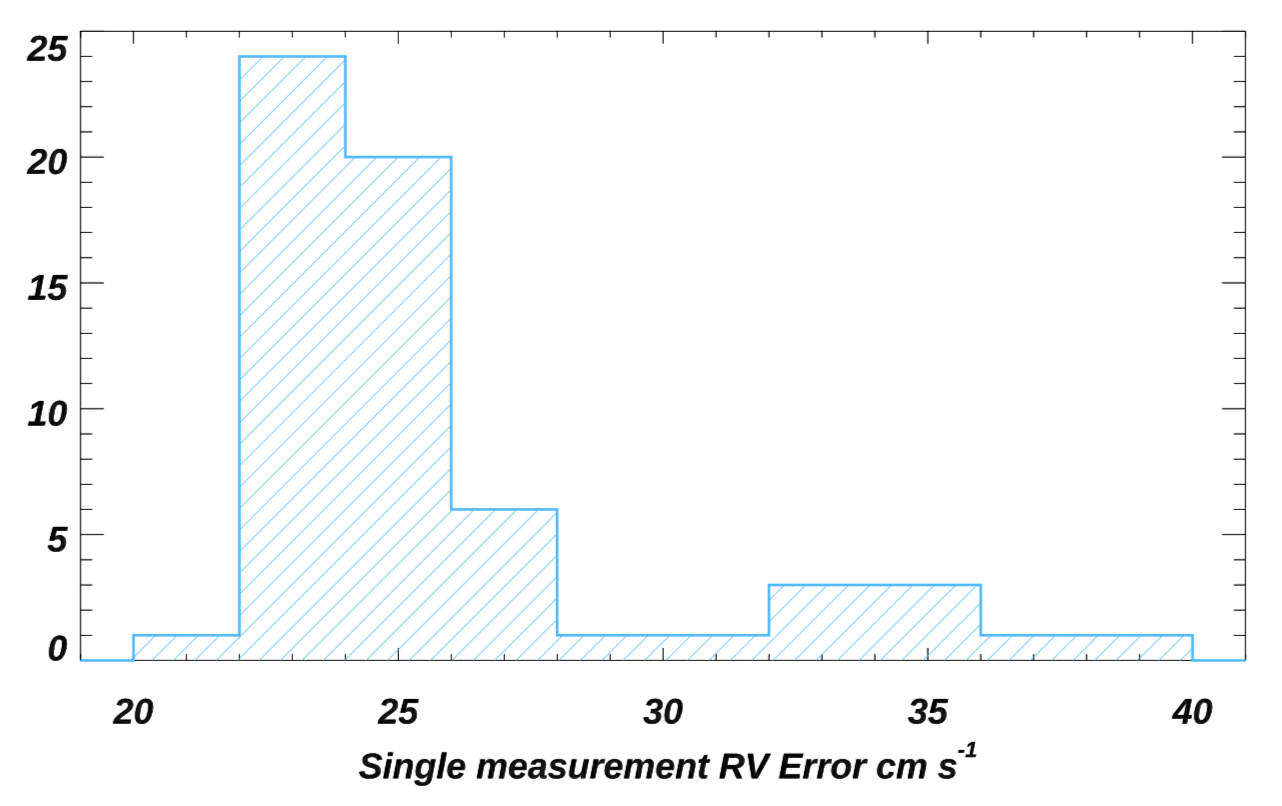}
    \caption{The distribution of single measurement errors for HD~3651 with EXPRES spectra obtained between August 2019 and February 2020. These unbinned measurement uncertainties are typically about 25 \cms.}
    \label{fig:smp}
\end{figure}

\subsection{Observations}

We obtained 61 EXPRES RV measurements between August 2019 and February 2020, which are presented in Table \ref{tab:3651vels}. The velocities were derived from optimally extracted spectra with forward modeling, as described in \citet{Petersburg_2020}. A histogram showing the distribution of tabulated single measurement errors is shown in Figure \ref{fig:smp}. The relatively high cadence of the \hundredearths\ allowed for excellent phase coverage over two orbital periods for HD~3651~b. On a night when we did not have telescope time, Lowell astronomer Maxime Devogele kindly yielded about 20 minutes of his time so that we could obtain a set of four spectra that allowed us to catch the rapid velocity change during periastron passage in November 2019.

\begin{deluxetable}{ccc}
\tablecaption{EXPRES  RVs of HD~3651 
\label{tab:3651vels}}
\tablewidth{0pt}
\tablehead{
\colhead{BMJD} & \colhead{Vel \ms}  & \colhead{Err \ms} }
\startdata
    18714.48211009 & -10.090 &   0.329  \\
    18714.49031062 &  -9.811 &   0.334  \\
    18715.47687595 & -11.452 &   0.377  \\
    18715.48546612 & -11.262 &   0.396  \\
    18716.41734766 & -12.623 &   0.311  \\
                 & \(\vdots\)  &            \\
\enddata
\tablecomments{The full data set is available online}
\end{deluxetable}

\subsection{Keplerian Fitting}
\begin{figure*}
    \centering
    \includegraphics[width=\linewidth]{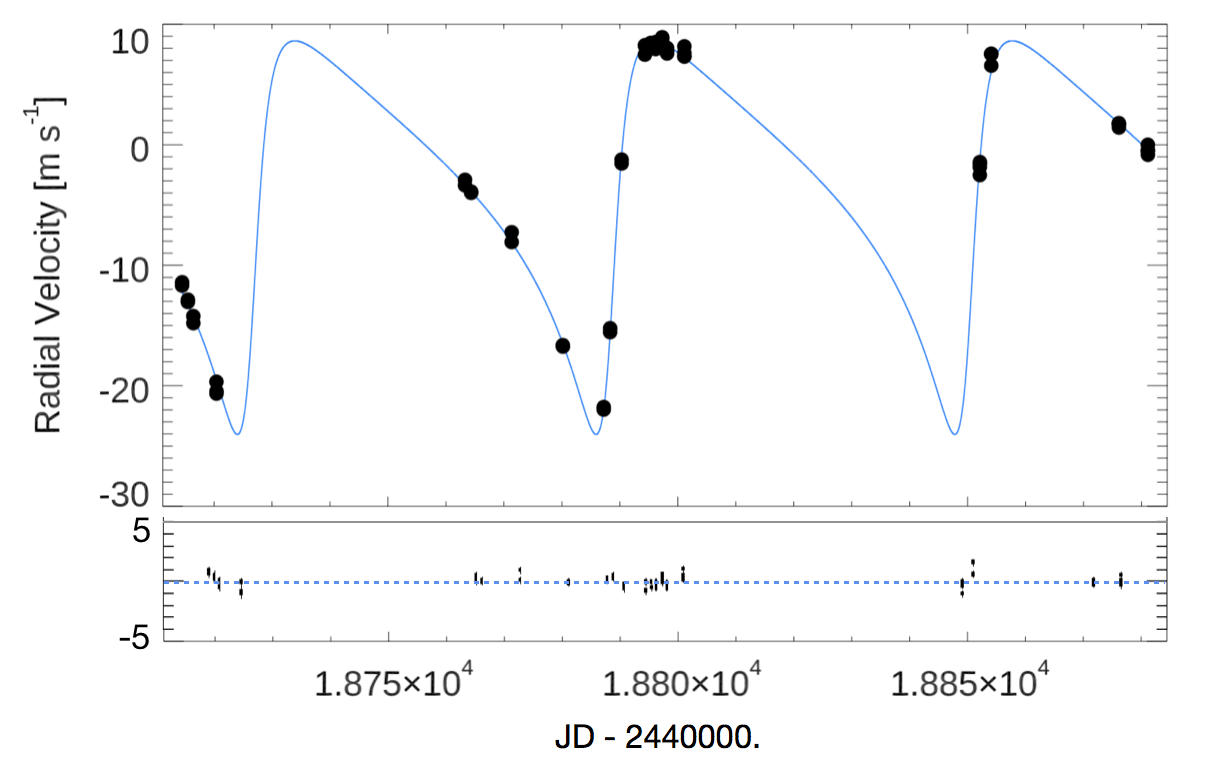}
    \caption{The time series radial velocity measurements of HD~3651 are fitted with a Keplerian model (shown with the blue curve). The residual velocities to this fit have an RMS of 58 \cms.}
    \label{fig:rv_timeseries}
\end{figure*}

Keplerian modeling of the velocities for HD~3651 was carried out using a Levenburg-Marquardt algorithm to fit the linearized Keplerian equations \citep{Wright2009} that are built into the IDL widget Keplerian Fitting Made Easy (KFME) developed by \citet{Giguere2012}. The best-fit model yields an orbital period of $61.88 \pm 0.55$ d, consistent with the better constrained orbital period of $62.26 \pm 0.075$ d modeled with the 17-year time baseline of Keck HIRES data \citep{butler_hires_2017}. Fixing the orbital period to $62.26$ d gives an equally good fit in the EXPRES data (Figure \ref{fig:rv_timeseries}) with a residual velocity RMS of 58 \cms.  There is no apparent periodicity in the residuals and a Lomb-Scargle periodogram shows no significant peaks.

To obtain uncertainties in the orbital parameters, we ran 1000 bootstrap Monte Carlo (MC) trials. For each MC trial, we fit a Keplerian orbit to the data, subtracted the best-fit model, scrambled the residuals (seeding with a random number generator), and added the scrambled residuals back to the best-fit model Keplerian velocities. We also carried out 1000 bootstrap MC trials with the Keck HIRES data \citet{butler_hires_2017}. The bootstrap MC errors on the orbital period and the time of periastron passage were smaller with the Keck HIRES data set because of the longer time baseline; however, the errors on all other orbital parameters were somewhat larger with the Keck HIRES data because of the larger error bars on those RV measurements. The independently fit model parameters for HD~3651~b using EXPRES and HIRES data are summarized in Table \ref{tab:3651_orb}.

\begin{deluxetable}{lll}
\tablecaption{Keplerian Model for HD~3651 b\label{tab:3651_orb}}
\tablewidth{0pt}
\tablehead{
\colhead{Parameter} & \colhead{EXPRES}  & \colhead{Keck HIRES} }
\decimalcolnumbers
\startdata
$P \, [d]$      & $61.88 \pm 0.55$   & $62.26 \pm 0.075$  \\
$T_p \, [d]$    & $58726.2 \pm 1.2 $ & $58726.68 \pm 0.5$ \\
$e$             & $0.606 \pm 0.09$    & $0.612 \pm 0.12$   \\
$\omega$        & $243.8 \pm 23.4$   & $231.9 \pm 41$     \\
$K  [\ms]$      & $16.93 \pm 0.22 $  & $17.15 \pm 0.9 $   \\
$M \sin i \, [M_\oplus]$ & $69.04 \pm 4.1$  & $66.88 \pm 5.9$\\
$a_{rel} [AU] $ & $0.284 \pm 0.002$  & $0.285 \pm 0.001$  \\
${\rm RMS} \ [\ms]$     & 0.58       & 3.4                \\
\enddata
\end{deluxetable}

\begin{figure}[htbp]
    \centering
    \includegraphics[width=\linewidth]{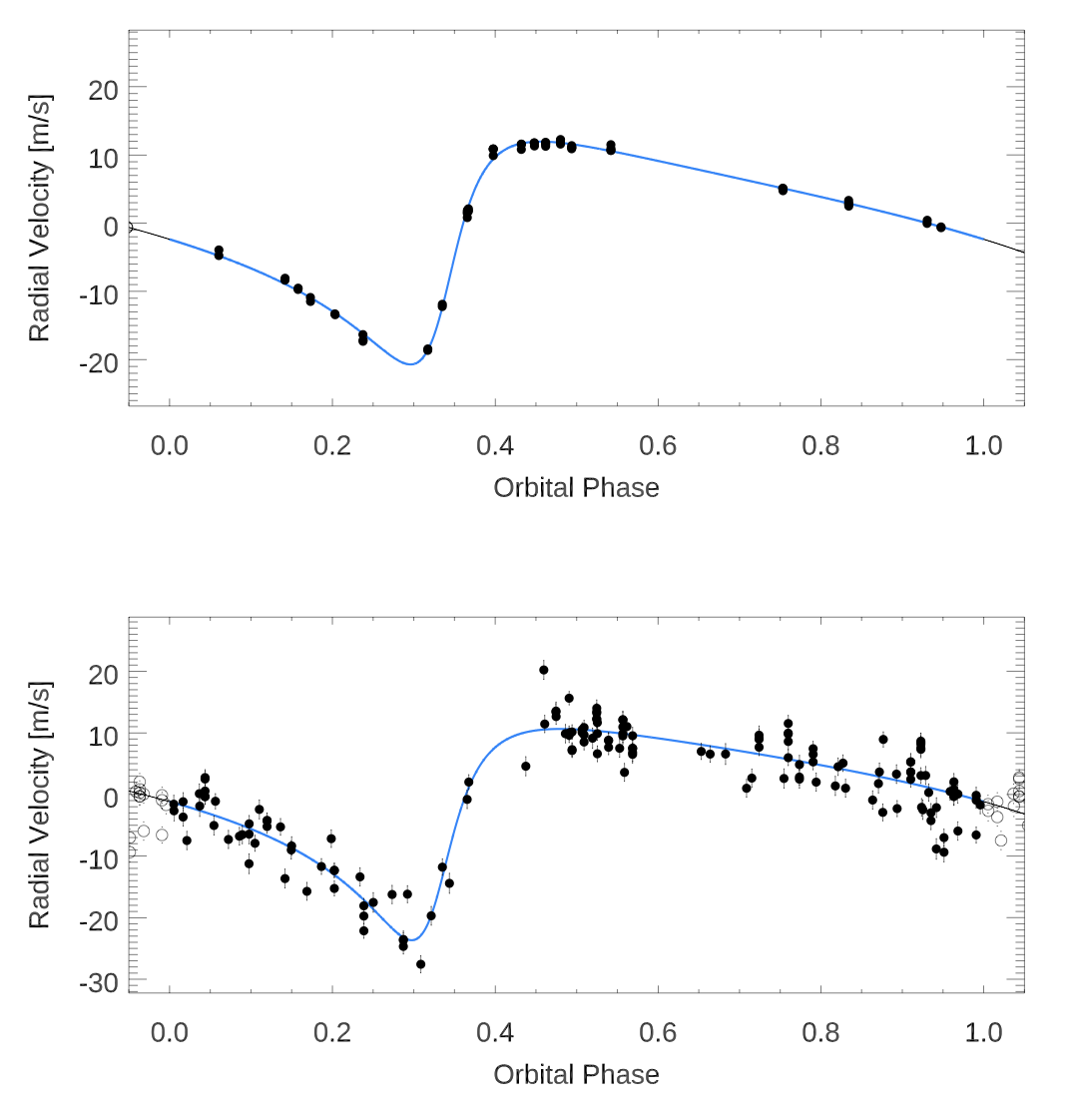}
    \caption{Phased radial velocities, Keplerian orbital fits, and residuals for  observations of HD~3651 b  obtained with EXPRES (top) and Keck archival data (bottom) from \citet{butler_hires_2017}. The RMS to the residuals for the EXPRES data is 0.58 \ms, compared to 3.46 \ms\ RMS for the Keck HIRES data.  }
    \label{fig:3651b}
\end{figure}

The phased orbital fit for the EXPRES velocities is shown in Figure \ref{fig:3651b} where we also include the fit to the archival Keck data phased and plotted in the same way for comparison. The EXPRES data show much less scatter in the RVs with scatter in the residuals that is about six times smaller than the Keck HIRES data. 

\section{APT Photometric Observations} \label{sec:3651_photometry}

To analyze the variability and look for serendipitous transits of the planet, we present 1192 photometric observations of HD~3651 acquired over an interval of 25 years from the 1993-94 to the 2017-18 observing seasons with the T4 0.75~m automatic photoelectric telescope (APT) at Fairborn Observatory in southern Arizona.  Our observations include the 10 observing seasons presented in the HD~3651b discovery paper of \cite{fischer_2003}. The T4 APT is equipped with a single channel photometer that uses an EMI 9124QB bi-alkali photomultiplier tube to measure the difference in brightness between the program star and three nearby comparison stars in the Str\"omgren $b$ and $y$ passbands.  To improve the photometric precision, we combine the differential $b$ and $y$ magnitudes into a single $(b+y)/2$ "passband".  The precision of a single observation with T4, as measured from pairs of constant comparison stars, is around ~0.0015~mag on good nights.  The T4 APT is described in \citet{henry_1999}, where further details of the telescope, precision photometer, and observing and data reduction procedures can be found.

Table \ref{tab:3651_phot} gives a summary of the photometric results of HD 3651.  We computed the differential magnitudes in the sense HD~3651 minus HD~3690, the best of our three comparison stars.  All magnitudes in the table refer to the average $(b+y)/2$ passband.  The standard deviations of a single observation from the seasonal means, given in column 4, range from 0.00105 to 0.00197 mag, so the night-to-night scatter in the observations is similar to the typical measurement uncertainty.  Period analysis found no significant variability within any observing season.  However, the seasonal means given in column 5 exhibit a range of $\sim0.003$~mag.  HD~3651 is a quiet star with a very low value of $\log R^{\prime}_{\rm HK} = -5.01$ and so probably exhibits no spot variability measurable at our precision.  The K0~Iab comparison star is the likely source of the long-term variability.

\subsection{APT Photometric Analysis}

\begin{figure}
\centering
    \includegraphics[width=\linewidth]{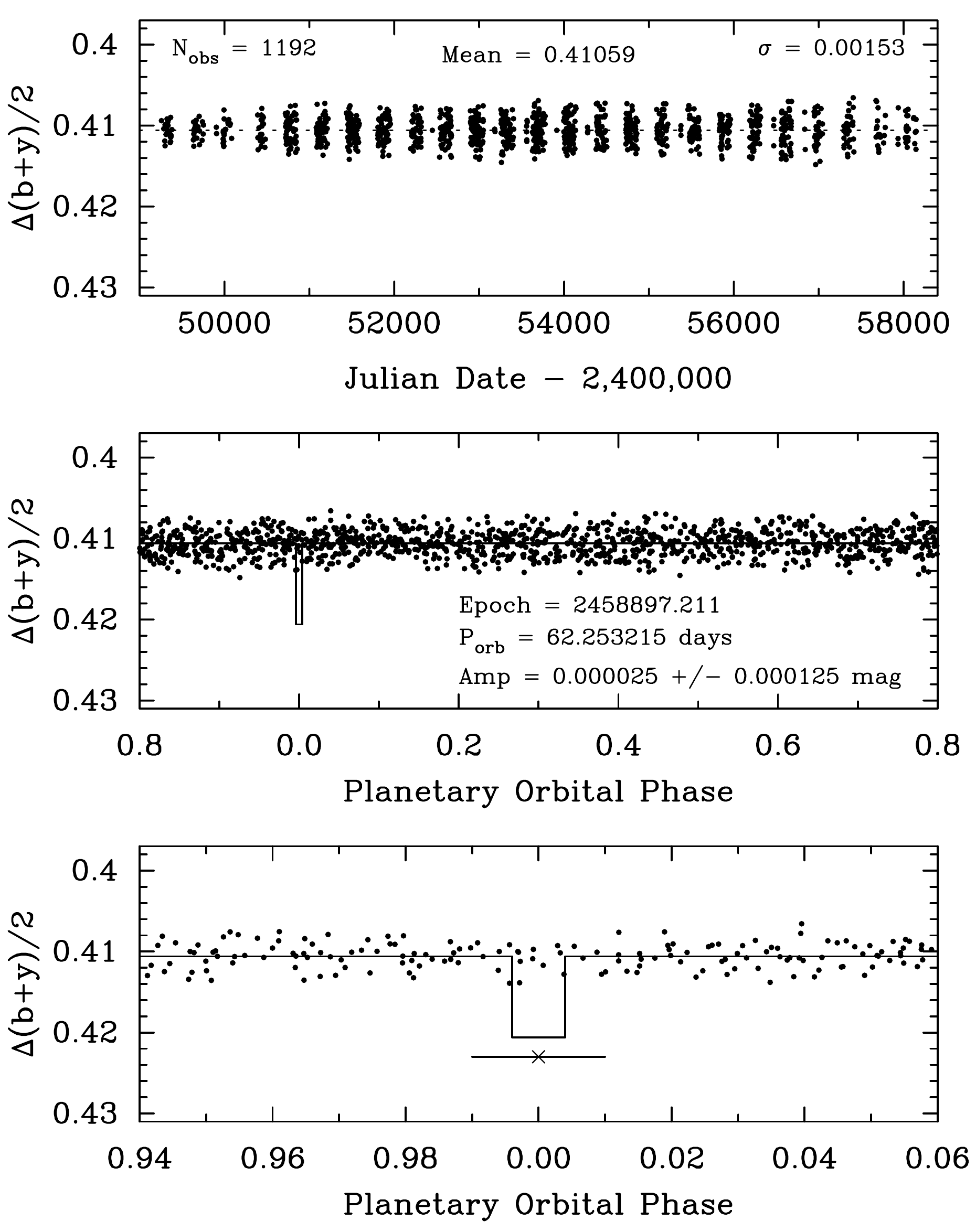}
\caption{Photometric observations of HD~3651 acquired over 25 years with the T4 0.75m APT at Fairborn Observatory.  $(Top)$:  Normalized observations show a scatter of 0.00153~mag from the mean, consistent with the measurement precision of the T4 APT.  $(Middle)$:  Normalized observations phased with the orbital period of HD~3651~b show an upper limit of any variability on the orbital period of 25~ppm.  $(Bottom)$:  The normalized observations around the predicted phase of transit show no evidence for a transit.  The point below the transit model shows the predicted time of transit center and its 1-sigma uncertainty.
\label{fig:3651_3panel}}
\end{figure}

For further analysis, the 25 observing seasons of APT photometry are normalized such that all 25 seasons have the same mean magnitude as the first.  This removes the long-term variability in the comparison star and in HD~3651, if any.  The 1192 normalized observations are plotted in the top panel of Figure~\ref{fig:3651_3panel}, where we note that the standard deviation of all observations from the normalized mean is 0.00153~mag, consistent with our measurement precision.  A period search of the complete normalized data set shows no significant variability between 1 and 100 days, as expected from HD 3651's low value of $\log R^{\prime}_{\rm HK}$ and the low scatter in the observations.  In particular, we find no signal in the vicinity of the estimated 44.5-day rotation period from \citet{fischer_2003}.

The observations are phased with the epoch of transit center and the orbital period of the planet and plotted in the middle panel of the figure.  A least-squares sine fit of the normalized observations to the planetary orbital period gives a peak-to-peak amplitude of $0.000025 \pm 0.000125$ mag.  This extremely low limit to any variability on the planetary orbital period is strong confirmation that the observed Doppler shifts are due to the planetary reflex motion of HD~3651~b.

Finally, in the bottom panel of Figure \ref{fig:3651_3panel}, we show the photometric observations near the transit epoch predicted from the radial velocities.  Given the orbital period and orientation of the orbit, the transit probability is only about 1\%. The transit duration is computed from the orbital elements and properties of the star, while the transit depth is estimated to be around 0.01~mag.  The horizontal error bar below the transit window is the uncertainty in the time of transit. It is clear there remains no evidence for transits of HD~3651~b.

\input{Tables/3651_photometry.tex}

\section{Dynamical Clearing from HD~3651~b}\label{sec:3651_interior_planets}

\begin{figure}[htbp]
    \centering
    \includegraphics[width=\linewidth]{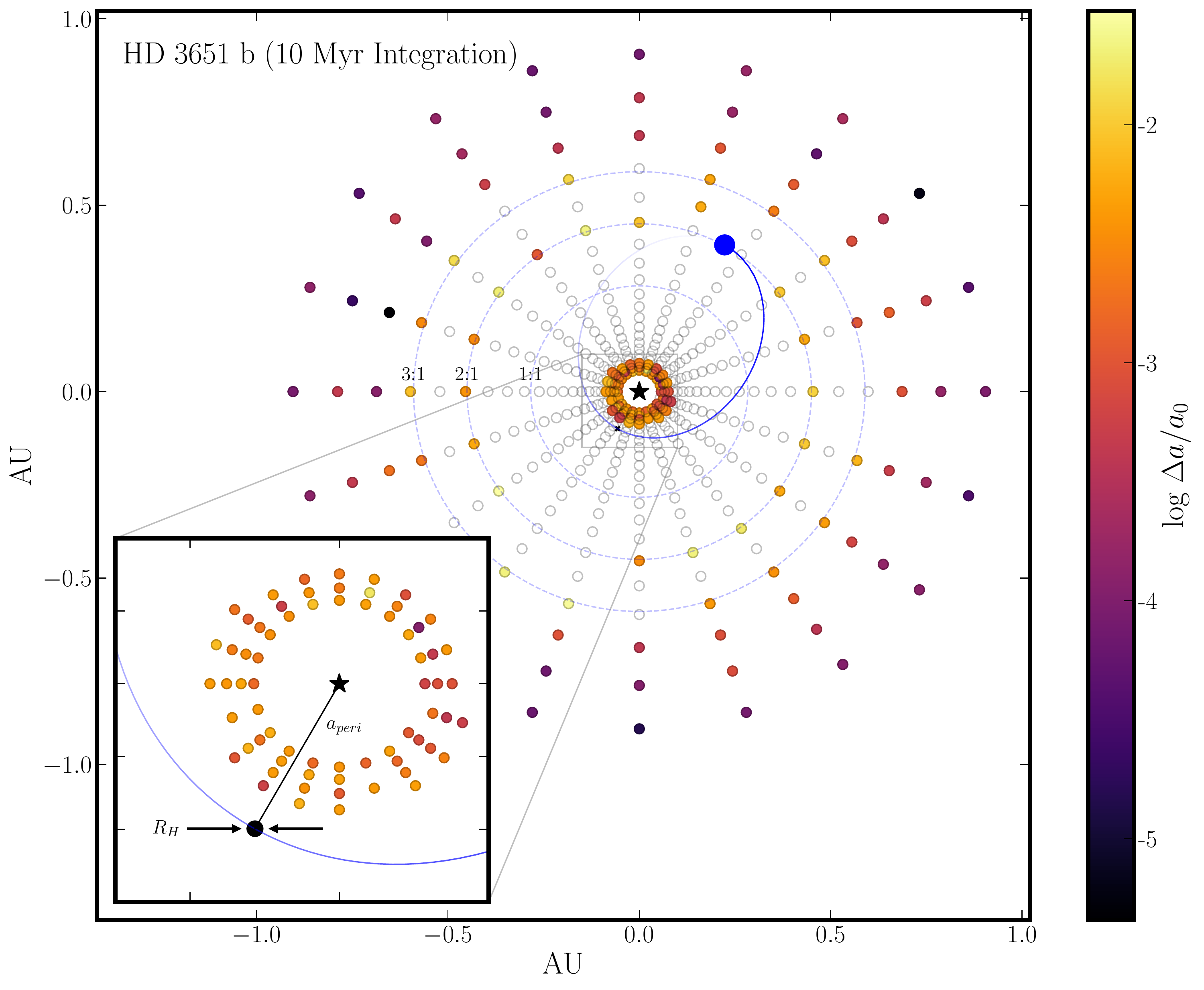}
    \caption{Orbital stability of zero-mass test particles placed at the starting locations shown. The color scale indicates the fractional change in semi-major axes over the length of the integration, with darker colors representing more stable initial orbits. The inset shows planets interior to HD~3651b that are able to survive for the entire simulation, with the planet's Hill radius shown for reference. All particles show some change in their orbital parameters over only 10 Myrs. Empty circles indicate initial positions of particles which suffered collision with the planet or star, or ejection from the system. We also mark circular orbits with 1:1, 2:1, and 3:1 mean motion resonance with HD~3651~b.}
    \label{fig:3651_stability}
\end{figure}

From the photometric observations, we can see that HD~3651 is a chromospherically inactive star.  The high eccentricity planet also makes it unlikely that there are interior planets that might contribute to additional low-amplitude, high-frequency noise in radial velocity fitting. We test the stability of various orbits using N-body simulations of the HD~3651~b system. We use the \texttt{MERCURIUS} hybrid symplectic integrator included in the \texttt{REBOUND} package \citep{Rein2019}. We circularly distributed zero-mass test particles throughout the HD~3651 system at 21 logarithmically spaced values of semi-major axis. Orbits ranged from $(1 - e)/2$ to $2(1 + e)$ times the semi-major axis of HD~3651~b. The simulation was integrated with 1-day timesteps for 10 Myr, with post-Newtonian precessional effects \citep{Nobili1986, Tamayo2020} taken into account. Particles which collided with the planet or host star were removed from the simulation. The outcomes are depicted in Figure~\ref{fig:3651_stability}. HD~3651~b clears out most particles within and slightly outside of its orbit, as expected from stability theory \citep[e.g.][]{Gladman1993}. Exceptions tend to be in mean motion resonance with the planet, but their orbits are significantly disturbed.  Although some particles survive within $\sim 0.09$ AU, their orbits are  disturbed by the planet as seen by fractional changes in semi-major axis. Remaining particles may not last on longer timescales; indeed, a few particles within this boundary were cleared out within 10 Myr. If a close-in Earth-mass planet could survive, it would leave a semi-amplitude signature of no more than $\sim 34$ \cms. This value is above our measurement precision, however, our sampling might not be sufficient for such a planet to be detectable in the periodogram. More massive or closer-in planets would leave more significant signals. 

\subsection{Detectability Limits}

Given the potential for at least marginally stable orbits at very short periods, we ran simulations to place mass limits on planets that might remain undetected with the data at hand.  From the dynamics, planets within 0.09 AU of the host star, or with less than 11-day orbits, may survive.  We therefore simulate possible planets out to 15-day periods and up to 15 Earth-masses.

For each planet, we simulate Keplerian RVs with identical temporal sampling as the observed data sets, preserving any window functions in the observations.  Representative white and red noise is added, scaled to the RMS of each data set's planet fit and the calculated activity level of HD 3651 respectively.  For each simulated planet, 1500 independent realizations of noise are generated.  We then compare the periodogram power of just the noise against the periodogram power of the noise with the injected RV signal at the injected period.

\begin{figure*}
    \centering
    \includegraphics[width=\linewidth]{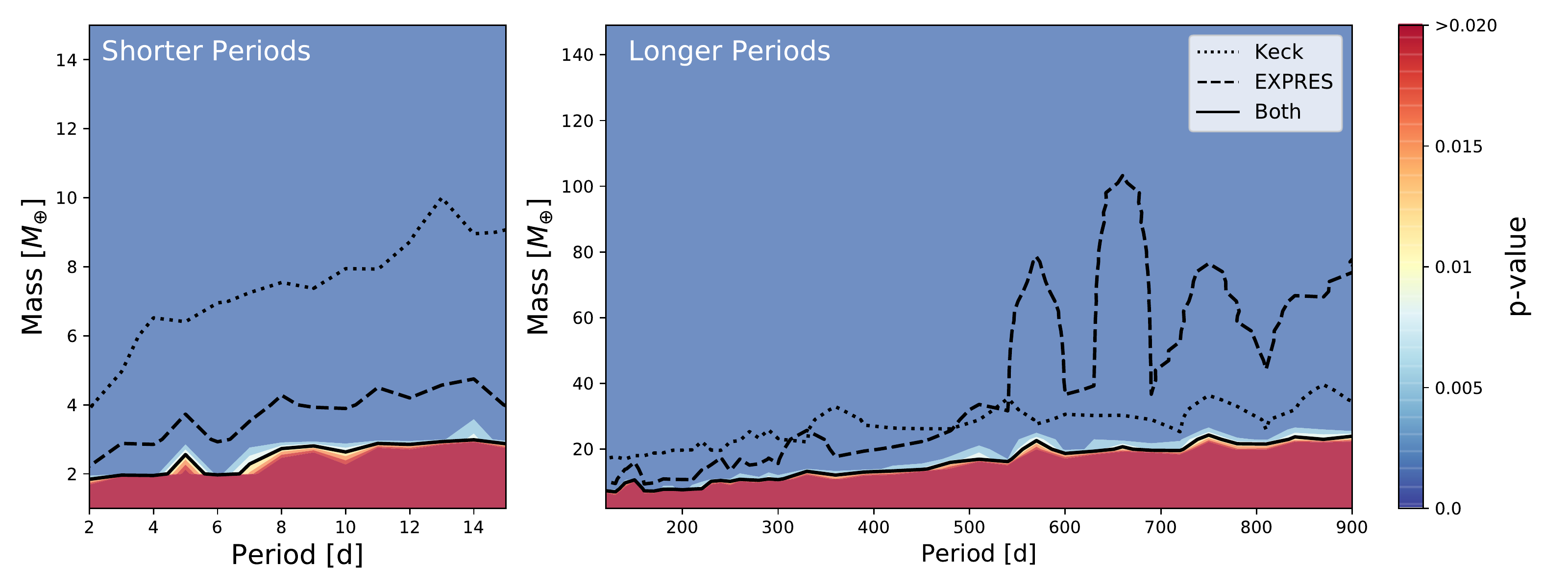}
    \caption{P-value contours with respect to mass and period showing the significance at which a planet would have been detected using the Keck HIRES and EXPRES data separately and together.  With a p-value of less than 0.01 considered a successful detection, the black solid line marks the border of detectability.  The detectability border when using just HIRES data or just EXPRES data is shown as dotted and dashed lines, respectively.  The left plot shows shorter periods while the right plot investigates longer periods.  Planets at the intervening periods would have been cleared out by HD 3651b's orbit (see Figure \ref{fig:3651_stability})}
    \label{fig:3651_detectability}
\end{figure*}

The p-value for each simulated planet gives the probability that the noise has equally or more significant signal as the injected signal plus noise, meaning the injected signal was buried in the noise.  A p-value of less than 0.01 is deemed a successful detection.  The results are shown in Figure \ref{fig:3651_detectability}, where contours are drawn in p-value space.  Blue areas indicate that if such a planet existed, it would have already been detected.  Only planets smaller than two or three Earth masses may still be hidden in the data.

Although we are most concerned about short-period signals, our dynamical simulations also showed possible stable orbits beyond the 2:1 mean motion resonance of HD~3651~b.  We performed the same detectability simulations as above for periods between 120 and 900 days with planet masses of $2 M_\oplus \leq M \leq 150 M_\oplus$ (see Figure \ref{fig:3651_stability} right). We find that planets with masses above 6~\mearth are excluded out to 160 days, $M > 8$ \mearth is excluded out to 210 days, and $M > 10$ is excluded out to 300 days.  There is a window near 150 days where a lack of the higher cadence EXPRES data would permit planets up to 10 \mearth to remain undetected.  All planets above 25-\mearth are excluded out to 900-day orbits.  The higher precision of EXPRES allows the EXPRES data to exclude more short-period planets with lower msses than even the 20-year time baseline of HIRES data.  However, as expected, the long time baseline of Keck data surpasses the 4-month time baseline of EXPRES data for orbital periods longer than a year.

\section{Conclusions}

The EXPRES \hundredearths\ has completed its first year of science operations. The instrumental precision of EXPRES has been measured to be 10 \cms\ \citep{Blackman_2020} and the single measurement precision on stars is about 30 \cms\ for spectra with \sn\ of 250 per pixel near 550~nm \citep{Petersburg_2020}. In addition to instrumental errors, we expect that other contributors to the RV error budget will include photospheric velocities, undetected low-mass planets in short-period orbits, telluric contamination, and errors from our analysis methods.

To eliminate some of the possible terms in the EPRV error budget we have used a new benchmark: 
the chromospherically quiet HD~3651 star, which hosts an eccentric, Saturn-mass planet in a $\sim 62$d orbit. We carried out N-body simulations to demonstrate that planets interior to the Saturn-mass planet, HD~3651~b, would be dynamically unstable; this eliminates undetected short-period, low-mass planets as one possible source of RV scatter in our data. The RMS scatter after fitting a Keplerian model is 58 \cms\ over $\sim 6$ months.  This suggests that added (roughly in quadrature) to our single measurement errors of 30 \cms, the remaining error terms (stellar activity, imperfectly modeled telluric contamination, and long-term instrumental drifts) contribute no more than about 50 \cms.  The residuals to this single planet fit give a good measure of the true long-term RV precision for chromospherically quiet stars observed with EXPRES. If similarly quiet stars exhibit more than 50 \cms\ RV scatter, then undetected, short-period planets are a good candidate for those RV variations.  

Importantly, the result that the intrinsic long term precision for chromospherically quiet stars is $\sim 50$ \cms\ helps to answer the question of whether RV precision in previous-era spectrographs was limited by the instrument, the analysis methods, or stability of the stellar photosphere. Several astronomers have long argued that photospheric velocities were the tall pole in the RV error budget. However, this work shows that in the case of quiet stars, the photospheric velocities were not dominating the error budget and this is validation for our decade-long effort to design EXPRES as a next generation EPRV spectrograph. While it is certainly true that photospheric velocities from active stars will be a strong contributor to the RV error budget, it is also likely that the high-fidelity data from next generation EPRV spectrographs offer the best chance for ultimately disentangling those photospheric velocities. The EPRV spectrographs will help the community to take the next big step along the path toward detecting smaller amplitude signals. This is a new parameter space for RV surveys. The statistical results from the \kepler mission suggest that this new parameter space will be a rich source of previously undetected exoplanets. 

EXPRES is not the only new EPRV instrument, and users of other EPRV spectrographs will want to evaluate their on-sky precision to track down instrumental issues that may affect their planet detection capability.  HD~3651~b is an ideal bench mark for demonstrating long term precision and for showing improvements relative to EXPRES. At a moderate northern declination, the star is observable by all current EPRV spectrographs and its brightness makes it an ideal standard for comparing RV precision and instrumental stability.

\acknowledgements{These results made use of the Lowell Discovery Telescope at Lowell Observatory. Lowell is a private, non-profit institution dedicated to astrophysical research and public appreciation of astronomy and operates the LDT in partnership with Boston University, the University of Maryland, the University of Toledo, Northern Arizona University and Yale University. We thank Lowell astronomer, Maxime Devogele, for generously yielding telescope time so that we could obtain the radial velocity point for HD~3651~b near periastron passage. We gratefully acknowledge ongoing support for telescope time from Yale University, the Heising-Simons Foundation, and an anonymous donor in the Yale community. We especially thank the NSF for funding that allowed for precise wavelength calibration and software pipelines through NSF ATI-1509436 and NSF AST-1616086 and for the construction of EXPRES through MRI-1429365. GWH acknowledges long-term support from NASA, NSF, Tennessee State University, and the State of Tennessee through its Centers of Excellence program. This material is based upon work supported by the National Science Foundation Graduate Research Fellowship under Grant No. DGE1122492 (RRP, LLZ, ABD).}

\bibliographystyle{aasjournal} 
\bibliography{main} 

\end{document}

%% file: authors.tex
\correspondingauthor{John M. Brewer} 
\email{jmbrewer@sfsu.edu}
\author[0000-0002-9873-1471]{John M. Brewer}

\affiliation{Department of Physics and Astronomy, San Francisco State University, 1600 Holloway Ave, San Francisco, CA 94132, USA}

\author[0000-0003-2221-0861]{Debra A. Fischer}
\author[0000-0002-0303-3276]{Ryan T. Blackman}
\author[0000-0001-9749-6150]{Samuel H. C. Cabot}
\author[0000-0002-5070-8395]{Allen B. Davis}
\author[0000-0002-3253-2621]{Gregory Laughlin}
\author[0000-0003-2369-0481]{Christopher Leet}
\author[0000-0001-7664-648X]{J. M. Joel Ong \chinesename}
\author[0000-0003-2168-0191]{Ryan R. Petersburg}
\author[0000-0002-4974-687X]{Andrew E. Szymkowiak}
\author[0000-0002-3852-3590]{Lily L. Zhao}
\affiliation{Department of Astronomy, Yale University, 52 Hillhouse Ave, New Haven, CT 06511, USA}

\author[0000-0003-4155-8513]{Gregory W. Henry}
\affil{Center of Excellence in Information Systems, Tennessee State University, \\ Nashville, TN 37209  USA}

\author[0000-0003-4450-0368]{Joe Llama}
\affiliation{Lowell Observatory, 1400 Mars Hill Rd, Flagstaff, AZ 86001, USA}

%% file: Tables/EXPRES_targets.tex
\begin{center}
\begin{longtable}{cccc}
\caption{Primary targets for the first phase EXPRES survey} \label{tab:targets} \\
\hline \multicolumn{1}{c}{\textbf{HD}} & \multicolumn{1}{c}{\textbf{Sp Type}} & \multicolumn{1}{c}{\textbf{Vmag}} &
\multicolumn{1}{c}{\textbf{P-mode [s]}} \\ \hline 
\endfirsthead
\multicolumn{4}{c}%
{{\bfseries \tablename\ \thetable{} -- continued from previous page}} \\
\hline \multicolumn{1}{c}{\textbf{HD}} & \multicolumn{1}{c}{\textbf{Sp Type}} & \multicolumn{1}{c}{\textbf{Vmag}} & 
\multicolumn{1}{c}{\textbf{P-mode [s]}} \\ \hline 
\endhead
\hline \multicolumn{4}{r}{{Continued...}} \\ \hline
\endfoot 
\hline
\endlastfoot
3651      & K0V    & 5.88  & 273          \\                     
4628      & K2V    & 5.74  & 236          \\
9407      & G6V    & 6.53  & 350          \\
10476     & K1V    & 5.24  & 260          \\
10700     & G8V    & 3.50  & 258          \\
16160     & K3V    & 5.5   & 220          \\
18803     & G8V    & 6.4   & 296          \\
19373     & G0V    & 3.8   & 340          \\
22049     & K0V    & 3.72  & 150          \\
26965     & K1V    & 4.43  & 258          \\
32147     & K3V    & 6.21  & 215          \\
34411     & G0V    & 4.8   & 492          \\
38858     & G4V    & 5.97  & 306          \\
50692     & G0V    & 5.75  & 365          \\
52711     & G4V    & 5.95  & 350          \\
55575     & G0V    & 5.6   & 428          \\
69830     & K0V    & 5.95  & 284          \\
71148     & G5V    & 6.3   & 406          \\
75732     & G8V    & 5.95  & 370          \\
76151     & G3V    & 6.0   & 330          \\
84737     & G2V    & 5.1   & 710          \\
86728     & G1V    & 5.4   & 433          \\
89269     & G5V    & 6.65  & 346          \\
95128     & G0V    & 5.04  & 450          \\
95735     & M2V    & 7.5   & 254          \\
99491     & K0V    & 6.5   & 200          \\
99492     & K3V    & 7.5   & 190          \\
101501    & G8V    & 5.34  & 260          \\
103095    & K0V    & 6.45  & 182          \\
104304    & K0V    & 5.55  & 396          \\
105631    & K0V    & 7.5   & 280          \\
110897    & G0V    & 5.95  & 290          \\
114783    & K0V    & 7.55  & 270          \\
115617    & G6.5V  & 4.74  & 316          \\
117043    & G6V    & 6.2   & 318          \\
122064    & K3V    & 6.52  & 200          \\
126053    & G1.5V  & 6.3   & 310          \\
127334    & G5V    & 6.36  & 472          \\
136923    & G9V    & 7.1   & 240          \\
141004    & G0V    & 4.42  & 540          \\
143761    & G0V    & 5.2   & 384          \\
146233    & G2V    & 5.5   & 420          \\
154345    & G8V    & 6.6   & 260          \\
157214    & G0V    & 5.39  & 220          \\
157347    & G5V    & 6.28  & 335          \\
158259    & G0V    & 6.5   & 222          \\
158633    & K0V    & 6.43  & 220          \\
159222    & G1V    & 6.4   & 300          \\
161797    & G5IV   & 3.4   & 700          \\
164922    & G9V    & 6.8   & 240          \\
166620    & K2V    & 6.4   & 255          \\
168009    & G2V    & 6.3   & 416          \\
182488    & G8V    & 6.36  & 325          \\
185144    & K0V    & 4.68  & 232          \\
186408    & G2V    & 5.95  & 465          \\
186427    & G5V    & 6.2   & 380          \\
190404    & K1V    & 7.3   & 220          \\
190406    & G1V    & 5.8   & 458          \\
191785    & K0V    & 7.3   & 220          \\
193664    & G3V    & 5.75  & 333          \\
197076    & G5V    & 6.44  & 340          \\
199960    & G1V    & 6.2   & 480          \\
210277    & K0V    & 8.57  & 397          \\
217014    & G2V    & 5.5   & 422          \\
218868    & K0V    & 7.0   & 330          \\
219134    & K3V    & 5.57  & 210          \\
221354    & K2V    & 6.76  & 285          \\ 
\end{longtable}
\end{center}

%% file: Tables/stellar_pars.tex
\begin{deluxetable}{ll}
\tablecaption{HD~3651 Parameters
\label{tab:pars}}
\tablewidth{0pt}
\tablehead{
\colhead{Parameter} & \colhead{Value}  }
\startdata
	Identifier & 54 Psc            \\
	           & HR~166            \\
	           & HIP~3093          \\
	V mag      & 5.88              \\
	B-V        & 0.85              \\
	dist [pc]  & 11.137 (0.007)    \\
	L          & 0.52              \\
	\mv        & 6.11              \\
	Sp Type    & K0V               \\
	Age [Gyr]  & 8.2 (3.0)         \\
	\teff [K]  & 5210 (30)         \\
	$\log g$   & 4.45 (0.15)       \\
	\feh       & 0.05 (0.05)       \\
	\vsini [\kms] &  1.7 (0.5)       \\
	Mass [\msun] & 0.8 (0.05)      \\
	Radius [\rsun] & 0.88 (0.02)   \\
	RV [\kms]    & -33.00 (0.16)     \\
	$\log R^\prime_{HK}$ & -5.01   \\
	$P_{rot}$ [d] & 44.5           \\
	\enddata
\end{deluxetable}

%% file: Tables/3651_photometry.tex
\begin{table}[ht!]
\footnotesize
\caption{APT Photometric Observations for HD~3651
\label{tab:3651_phot}}
\begin{tabular}{l|cccc}
\toprule
Obs     &     &  Date Range  & Sigma   & Seasonal Mean \\
Season        & $N_{obs}$ & HJD-2400000 & (mag) & (mag) \\
\hline
   1993--94   &  22 & 49258--49382 & 0.00105 & 0.41059(22) \\
   1994--95   &  22 & 49633--49750 & 0.00127 & 0.41078(27) \\
   1995--96   &  24 & 49904--50084 & 0.00124 & 0.41128(25) \\
   1996--97   &  22 & 50391--50480 & 0.00150 & 0.41170(32) \\
   1997--98   &  51 & 50718--50856 & 0.00153 & 0.41118(21) \\
   1998--99   &  61 & 51080--51218 & 0.00140 & 0.41111(17) \\
   1999--00   &  69 & 51434--51586 & 0.00141 & 0.41129(19) \\
   2000--01   &  57 & 51805--51952 & 0.00141 & 0.41129(19) \\
   2001--02   &  41 & 52193--52317 & 0.00149 & 0.41182(23) \\
   2002--03   &  52 & 52448--52673 & 0.00158 & 0.41147(22) \\
   2003--04   &  71 & 52897--53048 & 0.00150 & 0.41121(18) \\
   2004--05   &  69 & 53183--53405 & 0.00149 & 0.41063(18) \\
   2005--06   &  74 & 53557--53778 & 0.00163 & 0.41033(19) \\
   2006--07   &  70 & 53913--54140 & 0.00165 & 0.41122(20) \\
   2007--08   &  54 & 54275--54502 & 0.00173 & 0.40952(23) \\
   2008--09   &  63 & 54728--54865 & 0.00134 & 0.41040(17) \\
   2009--10   &  45 & 55091--55222 & 0.00171 & 0.41106(25) \\
   2010--11   &  55 & 55374--55595 & 0.00135 & 0.41110(18) \\
   2011--12   &  43 & 55830--55960 & 0.00144 & 0.41215(22) \\
   2012--13   &  56 & 56185--56315 & 0.00177 & 0.41201(24) \\
   2013--14   &  60 & 56468--56680 & 0.00181 & 0.41171(23) \\
   2014--15   &  35 & 56833--57043 & 0.00197 & 0.41127(33) \\
   2015--16   &  29 & 57296--57414 & 0.00190 & 0.41023(35) \\
   2016--17   &  20 & 57673--57784 & 0.00182 & 0.41006(41) \\
   2017--18   &  27 & 57933--58147 & 0.00159 & 0.40874(31) \\
\hline
\end{tabular}
\end{table}

%% file: main.bbl
\begin{thebibliography}{}
\expandafter\ifx\csname natexlab\endcsname\relax\def\natexlab#1{#1}\fi
\providecommand{\url}[1]{\href{#1}{#1}}
\providecommand{\dodoi}[1]{doi:~\href{http://doi.org/#1}{\nolinkurl{#1}}}
\providecommand{\doeprint}[1]{\href{http://ascl.net/#1}{\nolinkurl{http://ascl.net/#1}}}
\providecommand{\doarXiv}[1]{\href{https://arxiv.org/abs/#1}{\nolinkurl{https://arxiv.org/abs/#1}}}

\bibitem[{{Blackman} {et~al.}(2019){Blackman}, {Ong}, \&
  {Fischer}}]{blackman_measured_2019}
{Blackman}, R.~T., {Ong}, J.~M.~J., \& {Fischer}, D.~A. 2019, \aj, 158, 40,
  \dodoi{10.3847/1538-3881/ab24c3}

\bibitem[{Blackman {et~al.}(2017)Blackman, Szymkowiak, Fischer, \&
  Jurgenson}]{blackman_accounting_2017}
Blackman, R.~T., Szymkowiak, A.~E., Fischer, D.~A., \& Jurgenson, C.~A. 2017,
  The Astrophysical Journal, 837, 18, \dodoi{10.3847/1538-4357/aa5ead}

\bibitem[{{Blackman} {et~al.}(2020){Blackman}, {Fischer}, {Jurgenson},
  {Sawyer}, {McCracken}, {Szymkowiak}, {Petersburg}, {Ong}, {Brewer}, {Zhao},
  {Leet}, {Buchhave}, {Tronsgaard}, {Llama}, {Sawyer}, {Davis}, {Cabot},
  {Shao}, {Trahan}, {Nemati}, {Genoni}, {Pariani}, {Riva}, {Probst},
  {Holzwarth}, {Steinmetz}, {Fournier}, \& {Pawluczyk}}]{Blackman_2020}
{Blackman}, R.~T., {Fischer}, D.~A., {Jurgenson}, C.~A., {et~al.} 2020, arXiv
  e-prints, arXiv:2003.08852.
\newblock \doarXiv{2003.08852}

\bibitem[{Brems {et~al.}(2019)Brems, K{\"u}rster, Trifonov, Reffert, \&
  Quirrenbach}]{Brems:2019_rv_jitter}
Brems, S.~S., K{\"u}rster, M., Trifonov, T., Reffert, S., \& Quirrenbach, A.
  2019, Astronomy and Astrophysics, 632, A37

\bibitem[{Burke {et~al.}(2015)Burke, Christiansen, Mullally, Seader, Huber,
  Rowe, Coughlin, Thompson, Catanzarite, Clarke, Morton, Caldwell, Bryson,
  Haas, Batalha, Jenkins, Tenenbaum, Twicken, Li, Quintana, Barclay, Henze,
  Borucki, Howell, \& Still}]{Burke:2015_occur}
Burke, C.~J., Christiansen, J.~L., Mullally, F., {et~al.} 2015, The
  Astrophysical Journal, 809, 8

\bibitem[{{Butler} {et~al.}(2017){Butler}, {Vogt}, {Laughlin}, {Burt},
  {Rivera}, {Tuomi}, {Teske}, {Arriagada}, {Diaz}, {Holden}, \&
  {Keiser}}]{butler_hires_2017}
{Butler}, R.~P., {Vogt}, S.~S., {Laughlin}, G., {et~al.} 2017, \aj, 153, 208,
  \dodoi{10.3847/1538-3881/aa66ca}

\bibitem[{{Chaplin} {et~al.}(2019){Chaplin}, {Cegla}, {Watson}, {Davies}, \&
  {Ball}}]{chaplin_filtering_2019}
{Chaplin}, W.~J., {Cegla}, H.~M., {Watson}, C.~A., {Davies}, G.~R., \& {Ball},
  W.~H. 2019, \aj, 157, 163, \dodoi{10.3847/1538-3881/ab0c01}

\bibitem[{Cosentino {et~al.}(2014)Cosentino, Lovis, Pepe, Cameron, Latham,
  Molinari, Udry, Bezawada, Buchschacher, Figueira, Fleury, Ghedina, Glenday,
  Gonzalez, Guerra, Henry, Hughes, Maire, Motalebi, \&
  Phillips}]{Cosentino:2014_3651_harpsn}
Cosentino, R., Lovis, C., Pepe, F., {et~al.} 2014, in Proceedings of the SPIE,
  ed. S.~K. Ramsay, I.~S. McLean, \& H.~Takami, INAF - Telescopio Nazionale
  Galileo (Spain) (SPIE), 91478C

\bibitem[{DeGroff {et~al.}(2014)DeGroff, Levine, Bida, Cornelius, Collins,
  Dunham, Hardesty, Lacasse, Sweaton, Venetiou, Kermani, Massey, Foley, Larson,
  Sanborn, Strosahl, Winner, \& Pugh}]{DeGroff:2014_dct_performance}
DeGroff, W.~T., Levine, S.~E., Bida, T.~A., {et~al.} 2014, in Ground-based and
  Airborne Telescopes V, ed. L.~M. Stepp, R.~Gilmozzi, \& H.~J. Hall
  (International Society for Optics and Photonics), 91452C

\bibitem[{{Fischer} {et~al.}(2003){Fischer}, {Butler}, {Marcy}, {Vogt}, \&
  {Henry}}]{fischer_2003}
{Fischer}, D.~A., {Butler}, R.~P., {Marcy}, G.~W., {Vogt}, S.~S., \& {Henry},
  G.~W. 2003, \apj, 590, 1081, \dodoi{10.1086/375027}

\bibitem[{{Fischer} {et~al.}(2016){Fischer}, {Anglada-Escude}, {Arriagada},
  {Baluev}, {Bean}, {Bouchy}, {Buchhave}, {Carroll}, {Chakraborty}, {Crepp},
  {Dawson}, {Diddams}, {Dumusque}, {Eastman}, {Endl}, {Figueira}, {Ford},
  {Foreman-Mackey}, {Fournier}, {F{\H{u}}r{\'e}sz}, {Gaudi}, {Gregory},
  {Grundahl}, {Hatzes}, {H{\'e}brard}, {Herrero}, {Hogg}, {Howard}, {Johnson},
  {Jorden}, {Jurgenson}, {Latham}, {Laughlin}, {Loredo}, {Lovis}, {Mahadevan},
  {McCracken}, {Pepe}, {Perez}, {Phillips}, {Plavchan}, {Prato}, {Quirrenbach},
  {Reiners}, {Robertson}, {Santos}, {Sawyer}, {Segransan}, {Sozzetti},
  {Steinmetz}, {Szentgyorgyi}, {Udry}, {Valenti}, {Wang}, {Wittenmyer}, \&
  {Wright}}]{fischer_eprv_2016}
{Fischer}, D.~A., {Anglada-Escude}, G., {Arriagada}, P., {et~al.} 2016, \pasp,
  128, 066001, \dodoi{10.1088/1538-3873/128/964/066001}

\bibitem[{{Giguere} {et~al.}(2012){Giguere}, {Fischer}, {Howard}, {Johnson},
  {Henry}, {Wright}, {Marcy}, {Isaacson}, {Hou}, \& {Spronck}}]{Giguere2012}
{Giguere}, M.~J., {Fischer}, D.~A., {Howard}, A.~W., {et~al.} 2012, \apj, 744,
  4, \dodoi{10.1088/0004-637X/744/1/4}

\bibitem[{{Gladman}(1993)}]{Gladman1993}
{Gladman}, B. 1993, \icarus, 106, 247, \dodoi{10.1006/icar.1993.1169}

\bibitem[{{Henry}(1999)}]{henry_1999}
{Henry}, G.~W. 1999, \pasp, 111, 845, \dodoi{10.1086/316388}

\bibitem[{{Hoeijmakers} {et~al.}(2020){Hoeijmakers}, {Cabot}, {Zhao},
  {Buchhave}, {Tronsgaard}, {Kitzmann}, {Grimm}, {Cegla}, {Bourrier},
  {Ehrenreich}, {Heng}, {Lovis}, \& {Fischer}}]{Hoeijmakers_2020}
{Hoeijmakers}, H.~J., {Cabot}, S. H.~C., {Zhao}, L., {et~al.} 2020, arXiv
  e-prints, arXiv:2004.08415.
\newblock \doarXiv{2004.08415}

\bibitem[{Hsu {et~al.}(2018)Hsu, Ford, Ragozzine, \& Morehead}]{Hsu:2018_occur}
Hsu, D.~C., Ford, E.~B., Ragozzine, D., \& Morehead, R.~C. 2018, The
  Astronomical Journal, 155, 205

\bibitem[{Huang {et~al.}(2018)Huang, Liu, Chen, Zhang, Yuan, Xiang, Wang, \&
  Tian}]{Huang:2018_rv_standards}
Huang, Y., Liu, X.~W., Chen, B.~Q., {et~al.} 2018, The Astronomical Journal,
  156, 90

\bibitem[{{Isaacson} \& {Fischer}(2010)}]{isaacson_chromospheric_2010}
{Isaacson}, H., \& {Fischer}, D. 2010, \apj, 725, 875,
  \dodoi{10.1088/0004-637X/725/1/875}

\bibitem[{{Jurgenson} {et~al.}(2016){Jurgenson}, {Fischer}, {McCracken},
  {Sawyer}, {Szymkowiak}, {Davis}, {Muller}, \&
  {Santoro}}]{jurgenson_expres_2016}
{Jurgenson}, C., {Fischer}, D., {McCracken}, T., {et~al.} 2016, in \procspie,
  Vol. 9908, Ground-based and Airborne Instrumentation for Astronomy VI,
  99086T, \dodoi{10.1117/12.2233002}

\bibitem[{{Leet} {et~al.}(2019){Leet}, {Fischer}, \&
  {Valenti}}]{leet_tellurics_2019}
{Leet}, C., {Fischer}, D.~A., \& {Valenti}, J.~A. 2019, arXiv e-prints,
  arXiv:1903.08350.
\newblock \doarXiv{1903.08350}

\bibitem[{Levine {et~al.}(2012)Levine, Bida, Chylek, Collins, DeGroff, Dunham,
  Lotz, Venetiou, \& Kermani}]{Levine:2012_dct_commision}
Levine, S.~E., Bida, T.~A., Chylek, T., {et~al.} 2012, in Ground-based and
  Airborne Telescopes IV, ed. L.~M. Stepp, R.~Gilmozzi, \& H.~J. Hall
  (International Society for Optics and Photonics), 844419

\bibitem[{{Mayor} {et~al.}(2003){Mayor}, {Pepe}, {Queloz}, {Bouchy},
  {Rupprecht}, {Lo Curto}, {Avila}, {Benz}, {Bertaux}, {Bonfils}, {Dall},
  {Dekker}, {Delabre}, {Eckert}, {Fleury}, {Gilliotte}, {Gojak}, {Guzman},
  {Kohler}, {Lizon}, {Longinotti}, {Lovis}, {Megevand}, {Pasquini}, {Reyes},
  {Sivan}, {Sosnowska}, {Soto}, {Udry}, {van Kesteren}, {Weber}, \&
  {Weilenmann}}]{mayor_2003}
{Mayor}, M., {Pepe}, F., {Queloz}, D., {et~al.} 2003, The Messenger, 114, 20

\bibitem[{{Nobili} \& {Roxburgh}(1986)}]{Nobili1986}
{Nobili}, A., \& {Roxburgh}, I.~W. 1986, in IAU Symposium, Vol. 114, Relativity
  in Celestial Mechanics and Astrometry. High Precision Dynamical Theories and
  Observational Verifications, ed. J.~{Kovalevsky} \& V.~A. {Brumberg}, 105

\bibitem[{{Petersburg} {et~al.}(2018){Petersburg}, {McCracken}, {Eggerman},
  {Jurgenson}, {Sawyer}, {Szymkowiak}, \& {Fischer}}]{petersburg_modal_2018}
{Petersburg}, R.~R., {McCracken}, T.~M., {Eggerman}, D., {et~al.} 2018, \apj,
  853, 181, \dodoi{10.3847/1538-4357/aaa487}

\bibitem[{{Petersburg} {et~al.}(2020){Petersburg}, {Ong}, {Zhao}, {Blackman},
  {Brewer}, {Buchhave}, {Cabot}, {Davis}, {Jurgenson}, {Leet}, {McCracken},
  {Sawyer}, {Sharov}, {Tronsgaard}, {Szymkowiak}, \&
  {Fischer}}]{Petersburg_2020}
{Petersburg}, R.~R., {Ong}, J.~M.~J., {Zhao}, L.~L., {et~al.} 2020, \aj, 159,
  187, \dodoi{10.3847/1538-3881/ab7e31}

\bibitem[{{Probst} {et~al.}(2016){Probst}, {Lo Curto}, {{\'A}vila},
  {Brucalassi}, {Canto Martins}, {de Castro Le{\~a}o}, {Esposito},
  {Gonz{\'a}lez Hern{\'a}ndez}, {Grupp}, {H{\"a}nsch}, {Holzwarth},
  {Kellermann}, {Kerber}, {Mandel}, {Manescau}, {Pasquini}, {Pozna}, {Rebolo},
  {Renan de Medeiros}, {Stark}, {Steinmetz}, {Su{\'a}rez Mascare{\~n}o},
  {Udem}, {Urrutia}, \& {Wu}}]{probst_relative_2016}
{Probst}, R.~A., {Lo Curto}, G., {{\'A}vila}, G., {et~al.} 2016, in \procspie,
  Vol. 9908, Ground-based and Airborne Instrumentation for Astronomy VI,
  990864, \dodoi{10.1117/12.2231434}

\bibitem[{{Rein} {et~al.}(2019){Rein}, {Hernandez}, {Tamayo}, {Brown},
  {Eckels}, {Holmes}, {Lau}, {Leblanc}, \& {Silburt}}]{Rein2019}
{Rein}, H., {Hernandez}, D.~M., {Tamayo}, D., {et~al.} 2019, \mnras, 485, 5490,
  \dodoi{10.1093/mnras/stz769}

\bibitem[{{Ricker} {et~al.}(2015){Ricker}, {Winn}, {Vanderspek}, {Latham},
  {Bakos}, {Bean}, {Berta-Thompson}, {Brown}, {Buchhave}, {Butler}, {Butler},
  {Chaplin}, {Charbonneau}, {Christensen-Dalsgaard}, {Clampin}, {Deming},
  {Doty}, {De Lee}, {Dressing}, {Dunham}, {Endl}, {Fressin}, {Ge}, {Henning},
  {Holman}, {Howard}, {Ida}, {Jenkins}, {Jernigan}, {Johnson}, {Kaltenegger},
  {Kawai}, {Kjeldsen}, {Laughlin}, {Levine}, {Lin}, {Lissauer}, {MacQueen},
  {Marcy}, {McCullough}, {Morton}, {Narita}, {Paegert}, {Palle}, {Pepe},
  {Pepper}, {Quirrenbach}, {Rinehart}, {Sasselov}, {Sato}, {Seager},
  {Sozzetti}, {Stassun}, {Sullivan}, {Szentgyorgyi}, {Torres}, {Udry}, \&
  {Villasenor}}]{Ricker_2015}
{Ricker}, G.~R., {Winn}, J.~N., {Vanderspek}, R., {et~al.} 2015, Journal of
  Astronomical Telescopes, Instruments, and Systems, 1, 014003,
  \dodoi{10.1117/1.JATIS.1.1.014003}

\bibitem[{Soubiran {et~al.}(2018)Soubiran, Jasniewicz, Chemin, Zurbach,
  Brouillet, Panuzzo, Sartoretti, Katz, Le~Campion, Marchal, Hestroffer,
  Th{\'e}venin, Crifo, Udry, Cropper, Seabroke, Viala, Benson, Blomme,
  Jean-Antoine, Huckle, Smith, Baker, Damerdji, Dolding, Fr{\'e}mat, Gosset,
  Guerrier, {Guy, L. P.}, Haigron, Jan{\ss}en, Plum, Fabre, Lasne, Pailler,
  Panem, Riclet, Royer, Tauran, Zwitter, Gueguen, \&
  Turon}]{Soubiran:2018_rv_standards}
Soubiran, C., Jasniewicz, G., Chemin, L., {et~al.} 2018, Astronomy and
  Astrophysics, 616, A7

\bibitem[{{Tamayo} {et~al.}(2020){Tamayo}, {Rein}, {Shi}, \& {Hernand
  ez}}]{Tamayo2020}
{Tamayo}, D., {Rein}, H., {Shi}, P., \& {Hernand ez}, D.~M. 2020, \mnras, 491,
  2885, \dodoi{10.1093/mnras/stz2870}

\bibitem[{{Winn} \& {Fabrycky}(2015)}]{winn_occurrence_2015}
{Winn}, J.~N., \& {Fabrycky}, D.~C. 2015, \araa, 53, 409,
  \dodoi{10.1146/annurev-astro-082214-122246}

\bibitem[{{Wright} \& {Howard}(2009)}]{Wright2009}
{Wright}, J.~T., \& {Howard}, A.~W. 2009, \apjs, 182, 205,
  \dodoi{10.1088/0067-0049/182/1/205}

\end{thebibliography}
